\documentstyle[12pt]{article}

\begin{document}

\begin{titlepage}

\begin{center}

\hfill IMAFF-95/1\\
\hfill hep-th/9501036\\

\vskip .7in

{\bf ON THE STRING INTERPRETATION OF THE
$t{\bar t}$-GEOMETRY}

\vskip .3in

C\'esar G\'omez and Esperanza L\'opez

\vskip .3in
{\em Instituto
de Matem\'aticas y F\'isica Fundamental,}

{\em Serrano 123, 28006 Madrid (Spain)}

\end{center}

\vskip .3in

\begin{center} {\bf Abstract}  \end{center}

We derive the $t{\bar t}$-equations for generic $N\!=\!2$
topological field theories as consistency conditions for the
contact term algebra of topological strings. A generalization of
the holomorphic anomaly equation, known for the critical
${\hat c}\!=\!3$ case, to arbitrary non critical topological
strings is presented. The interplay between the non-trivial
cohomology of the $b$-antighost, gravitational descendants and
$\bar t$-dependence is discussed. The physical picture emerging
from this study is that the $\bar t$ (background) dependence of
topological strings with non-trivial cohomology for the
$b$-antighost, is determined by gravitational descendants.

\end{titlepage}

\section{Introduction}

\vspace{3mm}

An important issue in string dynamics is certainly
the study of the geometry of the space of two dimensional
quantum field theories, a question which is intimately connected
with the main problem of string background independence \cite{W}.
In the simplest setting of topological string theory \cite{W1} a
partial understanding of the background independence problem
\cite{SZ} can be reach through the study of the holomorphic
anomaly \cite{BCOV,BCOV1}. This is in part due to the more
precise knowledge, we have in this case, of the geometry of the
"theory space", which turns out to be a sort of generalized $N=2$
special geometry \cite{SG}
known as topological-antitopological, $t\bar{t}$, fusion
\cite{CV}.

The general structure of this geometry is defined by a vector
bundle with the base space parametrizing the different couplings,
and fiber $V$, the BRST cohomology of the corresponding
topological field theory. The main ingredient in the
characterization of the $t\bar{t}$-geometry is the existence
in $V$ of a hermitean scalar product $\langle \; , \; \rangle$
such that
\begin{equation}
\{ Q, Q^{\ast} \} = H
\label{i1}
\end{equation}
for $Q^{\ast}$ the adjoint of the BRST charge $Q$, and $H$ the
hamiltonian. Equation (\ref{i1}) together with the nilpotency
$Q^{2} \!=\! Q^{\ast 2}\! =\!0$
defines the SUSY $N\!=\!2$ algebra or, in more
mathematical terms, a Hodge system. Denoting by $|i\rangle$ a
topological basis, i.e. the cohomology of $Q$, the
$t\bar{t}$-metric is defined by
\begin{equation}
\langle {\bar j} | i \rangle = g_{i {\bar j}}
\label{i2}
\end{equation}
The derivation of the $t\bar{t}$-geometry requires now to
introduce a connection by
\begin{equation}
\langle {\bar k} | D_i | j \rangle =0 \; , \hspace{1cm}
\langle {\bar k} | {\bar D}_{\bar i} | j \rangle =0
\label{i3}
\end{equation}
with respect to which the metric $g_{i {\bar j}}$
is covariantly constant.
The $t\bar{t}$-equations for this connection are
\begin{equation}
[ D_i , {\bar D}_{\bar j} ] = - [ C_i , {\bar C}_{\bar j} ]
\label{i4}
\end{equation}
with the $C's$ the ring structure constants.

\vspace{3mm}

{\it 1.1 Special geometry}.
In order to see the strong analogy with
special geometry let us consider the example of a Calabi Yau
3-fold $M$ where the
space of couplings is identified with the moduli of complex
structures, $X$. In this case the relevant BRST cohomology states
correspond to elements in $H^{2,1}(M)$. The role of the
vacuum, making possible the map from operators to states, is
played by the holomorphic top form $\Omega^{(3,0)}$ of $M$, with
the state-operator map being determined by the isomorphism
between $H^{0,1}(T M)$ and $H^{2,1}(M)$.
The full BRST cohomology is given by $H^3 (M) \!=\! H^{(3,0)}\!
\otimes \! H^{(2,1)} \! \otimes \! H^{(1,2)}\! \otimes \!H^{(0,3)}$,
whose elements will be denoted respectively as $\Omega$, $V_l$,
$V_{\bar l}$, ${\bar \Omega}$. The hermitean scalar product of
forms $\alpha$, $\beta \! \in \! H^3$ is
defined by means of the simplectic form on M
\begin{equation}
\langle \alpha , \beta \rangle = \int \alpha \wedge \beta
\label{sg1}
\end{equation}
and therefore the adjoint of an state $(p,q)$ is an element
$(3\!-\!p,3\!-\!q)$.

Infinitesimal motion on the moduli space of complex structures
mix $(p,q)$-forms with $(p\!\pm \! 1 , q\! \mp \! 1)$-forms. In
particular, the top form $\Omega$ mixes only with $(2,1)$-forms
\begin{equation}
\partial_i \Omega = V_i + a_i \Omega
\label{sg2}
\end{equation}
with $a_i$ certain function, not globally holomorphic.

Using now the definition (\ref{i3}) of covariant derivative
and the inner product (\ref{sg1}), we deduce that the projection
of $\partial_i V_j$ on $H^{(2,1)}$-forms defines the connection
on $X$
\begin{equation}
\partial_i V_j = A_{ij}^k + ...
\label{sg3}
\end{equation}
where by points we mean degree $(1,2)$ and $(3,0)$ contributions.
Special geometry allow to express the Yukawa couplings in the
following way
\begin{equation}
C_{ijk} = - \int \Omega \wedge \partial_i \partial_j \partial_k
\Omega =  \int \partial_i \Omega \wedge \partial_j \partial_k \Omega
\label{sg4}
\end{equation}
{}From this we easily get
\begin{equation}
\partial_j (\partial_k \Omega) = C_{jk}^{\bar l} V_{\bar l} +
(2,1) + (3,0)
\label{sg5}
\end{equation}
and therefore
\begin{equation}
\partial_i V_j = C_{ij}^{\bar l} V_{\bar l} + A_{ij}^k V_k + (3,0)
\label{sg6}
\end{equation}

Under an infinitesimal motion in the $\bar t$-direction on $X$,
and using (\ref{sg2})-(\ref{sg5}), equation (\ref{sg6}) give
raise to the curvature equations for the connection $A_{ij}^k$
\begin{equation}
\partial_{\bar l} A_{ij}^k = G_{j {\bar l}} \delta_{i}^k -
{\bar C}_{\bar l}^{kn} C_{ijn}
\label{sg7}
\end{equation}
where $G_{j {\bar l}}\!=\!\partial_{\bar l} a_j$
(see eq.(\ref{sg2})).
Equation (\ref{sg7}) is a particular case of the
general $t{\bar t}$-equation (\ref{i4}).

\vspace{3mm}

{\it 1.2 String representation}.
The string interpretation of the
$t\bar{t}$-geometry is based on the following general
philosophy. Given a topological field theory parametrized by the
couplings $(t_{i},\bar{t}_{i})$ the variation of the correlators
under small changes of the couplings is given by string
amplitudes of the corresponding topological matter theory
coupled to topological gravity. The reason for this is, of
course, that a variation of the couplings corresponds to
integrate over the world sheet the perturbing operator.
This is clear from the lagrangian representation of a perturbed
TFT
\begin{equation}
{\cal L} = {\cal L}_0 + \sum_i t_i \int \phi_{i}^{(2)} + \sum_{\bar
i} {\bar t}_{\bar i} \int {\bar \phi}_{\bar i}^{(2)}
\label{i5}
\end{equation}
where
\begin{equation}
\phi_{i}^{(2)} = \{ Q^* , [ {\bar Q}^* , \phi_{i} ] \} \; ,
\hspace{1cm} {\bar \phi}_{\bar i}^{(2)} = \{ Q , [ {\bar Q} ,
{\bar \phi}_{\bar i} ] \}
\label{i6}
\end{equation}
and $\phi_i$, ${\bar \phi}_{\bar i}$ are respectively chiral
and antichiral primary fields.
Therefore, the variation of correlators of the TFT under small
changes of couplings
implies the definition of a form on the moduli space
of the punctured Riemann surface, and this is string theory.
In this framework the $t\bar{t}$-connection on the space of
couplings should be determined by means of contact terms,
moreover the consistency conditions of the contact term algebra
should correspond to geometrical constraints. In
topological string theories \cite{VV} these
contact terms are computed by the cancel propagator argument and
they correspond to the contribution at the boundary of the
moduli space defined when two punctures collide. This kind of
computation strongly depends on the way the string measures on
the moduli space have been defined, and in particular on the type of
$(b,c)$ ghost system we use. In the case we are interested in
associating forms on the moduli of punctured Riemann surfaces
to infinitesimal changes of the couplings $(t_{i},\bar{t}_{i})$,
we are forced to use as the $b$ antighost the supercharge
$Q^{\ast}$ which is the adjoint of the BRST charge. The so defined
string differs from the standard bosonic string in a crucial
aspect, namely the cohomology of $b$ is now non trivial
\cite{BCOV1}. We will call Hodge strings those for which the $b$
antighost possess non trivial cohomology and such that the pair
$(Q,b)$ satisfies the usual Hodge relations\footnote{We thank
A.Losev for stressing to us the connection with Hodge theory}.

Summarizing, the string interpretation of the $t\bar{t}$-geometry
that we want to present in this paper will be based on
the following dictionary \cite{U}
\begin{eqnarray}
t{\bar t}-connection & \Longleftrightarrow & Contact \;\;\; Terms
\nonumber \\
t{\bar t}-equations & \Longleftrightarrow & Consistency \;\;\;
Conditions
\nonumber
\end{eqnarray}

\vspace{3mm}

{\it 1.3 Hodge equivariance and background independence}.
The
simplest consequence of the non trivial cohomology for the $b$
antighost is the absence, in the equivariant cohomology defined
by the Hodge pair $(Q,b)$, of gravitational descendants.
To define the physical states in this equivariant
cohomology is more than we need for constructing good string
amplitudes independent of the local world sheet
coordinates. Moreover these Hodge strings present a severe form
of BRST anomaly, namely the holomorphic anomaly \cite{BCOV,BCOV1},
which implies a $\bar{t}$-dependence
of the amplitudes. Taking into account that both, the appearance
of the holomorphic anomaly and the absence in "Hodge"
equivariance of gravitational descendants share a common origin,
namely the non trivial cohomology of the $b$ antighost,
it is natural to
try to connect them. A possible way to do it is trying to match the
$\bar{t}$-dependence with the contribution of gravitational
descendants.
In our approach this phenomena shows up in the form of new mixed
$t\bar t$-contact terms.

The plan of the paper is as follows. In section 2 we introduce
a contact term algebra whose consistency conditions imply the
whole set of $t\bar{t}$-equations. We
will discuss also what is the meaning of Hodge strings.
In section 3 we will propose a generalization of the
holomorphic anomaly, presented in \cite{BCOV1} for ${\hat
c}\!=\!3$ topological strings, to topological strings with
arbitrary $\hat c$.

\vspace{7mm}

\section{Contact Terms and $(t,\bar{t})$-Fusion}

\vspace{3mm}

In this section we will proceed to derive the $t{\bar
t}$-geometry from a contact term algebra. We will work in the
following general setting. Given a generic two dimensional
topological field theory, we consider the $(t{\bar t})$ space of
couplings defined by $N\!=\!2$ preserving perturbations. Our aim
will be to find the behaviour of the metric (2) under these
perturbations in terms of the contact terms of a topological
string.

\vspace{3mm}

\subsection{$t{\bar t}$-connection and Contact Terms}

\vspace{3mm}

The $t{\bar t}$-connection is defined in reference \cite{CV} by
the condition
\begin{equation}
\langle {\bar j} | D_i |k \rangle = 0
\label{co}
\end{equation}
This corresponds to the standard Levi-Civita definition of
connection, where the variation of the physical state
$|k\rangle$, induced by the $N\!=\!2$ preserving perturbation
$\delta t_i \int \phi_{i}^{(2)}$, is orthogonally projected, with
respect to the hermitean scalar product, on the basis of BRST
physical states. By a contact term representation of (\ref{co}),
we mean
\begin{equation}
\langle {\bar j} | \partial_i | k \rangle = \langle {\bar j} |
C(i,k) \rangle
\label{ct}
\end{equation}
with $C(i,k)$ a contact term defined in some topological
string theory.

Before entering into the explicit definition of these contact
terms, we will restrict them by imposing some consistency
conditions. These conditions will be motivated by the string
interpretation of these contact terms \cite{VV,DN,L}.
The string meaning of
$|C(i,k)\rangle$ can be symbolically represented as follows
\begin{equation}
|C(i,k)\rangle = \int \! i \; |k \rangle
\label{st}
\end{equation}
where we are thinking $|k\rangle$ as the state created by
inserting at the origin of the disk the field $\phi_k$, and
$\int i$ as the integration of $\phi_{i}^{(2)}$ in a
infinitesimal neighbourhood of the insertion point of $\phi_k$.
The consistency conditions are now determined by imposing
independence of the order of integration
\begin{equation}
\int i \int j \; |k\rangle = \int j \int i \; |k \rangle
\label{cc}
\end{equation}
The contribution to each term of (\ref{cc}) is given by
\begin{equation}
\int i \int j | k \rangle = \int c(j,i) |k\rangle + \int i
|C(j,k)\rangle
\label{ccs}
\end{equation}
where we introduce an explicit difference between
operator-operator contact terms $c(i,j)$ and the operator-state
contact terms $|C(i,j)\rangle$ defined by (\ref{st}).

The relation between $c(i,j)$ and $|C(i,j)\rangle$ will
reflects, of course, the standard state-operator relation or
equivalently the way we define a reference vacuum state. In
twisted topological field theories, the definition of the vacuum
requires to soak up fermionic zero modes and therefore requires
the insertion at the origin of the disk of some spectral flow like
operators. Denoting generically by $\Phi$ the operator used in
the definition of the vacuum, we can define the
operator-operator contact term $c(i,j)$ by the following
decomposition
\begin{equation}
|C(i,j)\rangle \equiv c(i,j) |\Phi\rangle + \phi_j |C(i,\Phi)\rangle
\label{cts}
\end{equation}
where the second term in the r.h.s. of (\ref{cts}) represents the
state obtained by inserting $\phi_j$ on the perturbed "vacuum".

At this point we should stress the difference between a marginal
perturbation, which preserves charge conservation, and a
massive perturbation.
When the $U(1)$ charge of the $N\!=\!2$ algebra is conserved,
the variation of the vacuum state is proportional to itself
\begin{equation}
|C(i, \Phi)\rangle = f_i |\Phi\rangle
\end{equation}
and in consequence the vacuum $|\Phi\rangle$ defines a line
subbundle, ${\cal L}$, over the moduli space of marginal
perturbations. In this case the $t{\bar t}$-geometry reduces to
special geometry and, in particular, the existence of $\cal
L$ translate into the existence of a K\"ahler potential from
which to derive all the relevant geometrical quantities. The
existence of $\cal L$ is
also important for the definition of covariant string
amplitudes, which will be discussed in the next section.

Under a generic massive perturbation breaking charge
conservation, the variation of the vacuum can have projection
into any harmonic state\footnote{Let us note that
\begin{equation}
|C(i,P)\rangle=|C(i,\Phi)\rangle
\end{equation}
where P is the puncture operator. This is due to $c(i,P)\!=\!0$
in any TFT.}
\begin{equation}
|C(i, \Phi)\rangle = A_{i0}^n |n\rangle
\end{equation}
and the notion of vacuum subbundle disappear.

In the context of Landau-Ginzburg theories \cite{DVV},
in which the
chiral fields are given in terms of the superpotential $W$ by
\begin{equation}
\phi_i  = \frac{\partial W}{\partial t_i}
\end{equation}
it is safe to assume for the operator-operator contact terms,
generically defined as
$c(i,j)\!\equiv\!\frac{\partial\phi_j}{\partial t_i}$,
the symmetry condition
\begin{equation}
c(i,j)=c(j,i)
\label{csy}
\end{equation}
We will suppose that this holds in general,
including all possible sources of asymmetric contributions to
the contact terms in the last piece of equation (\ref{cts}).
Notice that the analog of this asymmetry in pure topological
gravity \cite{VV} results from localizing the curvature at the
insertion points, with $\Phi$ defining there the curvature
insertion operator. In our formal definition of the contact term
algebra, we transfer the whole problem of the asymmetry of
contact terms into the general state-operator relation.

\vspace{3mm}

\subsection{$t{\bar t}$-geometry and Contact Term Algebra}

\vspace{3mm}

Our task now will be, assuming (\ref{ct}), to derive the $t{\bar
t}$-geometry \cite{CV} using only consistency conditions of
type (\ref{cc}). In order to do that, we need to introduce mixed
topological-antitopological contact terms. Our philosophy in
this section will be to introduce formally these contact terms
and only, after solving the consistency conditions and matching
the $t{\bar t}$-geometry, to look for a proper string
representation of these mixed contact terms. We will use for
operator-state contact terms notation (\ref{st}). Taking into
account that we are looking for the contact term algebra of a
topological string theory we will work, from the beginning, with
the whole tower of gravitational descendants,
$\sigma_n(i)$ with $n>0$ and $i$ running over the
chiral primary fields.
The topological
part of the contact term algebra is defined by
\begin{eqnarray}
&& \int i \; | j \rangle  =  A_{ij}^k | k \rangle
\label{tct} \\
&& \int \sigma_n (i) \; | \sigma_m (j) \rangle  =  A_{ij}^{k} |
\sigma_{n+m} (k) \rangle + C_{ij}^{l} | \sigma_{n+m-1} (l)
\rangle \; \; , \; \; \; n+m>0 \nonumber
\end{eqnarray}
with $C_{ij}^l$ and $A_{ij}^k$ some unknown tensors. From
(\ref{tct}) and (\ref{ct}) we observe that
the tensor $A_{ij}^k$ will play the role of the $t{\bar
t}$-connection. The second piece in (\ref{tct}.2) is the standard
one we expect for topological matter coupled to topological
gravity \cite{Li}. Now we complete (\ref{tct})
with the following mixed $t{\bar t}$-contact terms
\begin{eqnarray}
\int {\bar a} | \sigma_n (i) \rangle & = & H_{{\bar a} i}^l
|\sigma_{n+1} (l) \rangle \label{atct} \\
\int \sigma_n (i) | {\bar a} \rangle & = & {\widetilde H}_{i {\bar
a}}^l |\sigma_{n+1} (l) \rangle \nonumber
\end{eqnarray}
with the $\bar a$'s one to one related to the $\bar t$-$N\!=\!2$
preserving perturbations, and their associated states $|{\bar
a}\rangle$ to be determined in the process of solving the
consistency conditions\footnote{The states $|{\bar a}\rangle$
should not be confused with antitopological Ramond vacua.}.
The tensors $H_{{\bar a} i}^{l}$ and
${\widetilde H}_{i{\bar a}}^{l}$ are in principle different.
The main
feature of (\ref{atct}) is the appearance of gravitational
descendants (consider the case $n\!=\!0$) in the
topological-antitopological fusion. In fact, a natural way to
read equation (\ref{atct}), that we will discuss latter, is as a
procedure to associate with pure matter states their
gravitational descendants.

One more ingredient is still necessary before entering to solve
the consistency conditions (\ref{cc}). We will generalize the
contact terms (\ref{tct}) and (\ref{atct}) to the case in which
functions $f(t,{\bar t})$ are present, in the following way
\begin{eqnarray}
\int i \; ( f(t,{\bar t}) |A\rangle) & = & \partial_i f(t,{\bar
t}) |A\rangle + f(t,{\bar t}) | C(i,A) \rangle \nonumber \\
\int {\bar a} \; ( f(t,{\bar t}) |A\rangle) & = & \partial_{\bar a}
f(t,{\bar t}) |A\rangle + f(t,{\bar t}) | C({\bar a},A) \rangle
\label{d} \\
\int \sigma_n (i) \; ( f(t,{\bar t}) |A\rangle) & = & \partial_i
f(t,{\bar t})| \sigma_n (A) \rangle + f(t,{\bar t})
| C(\sigma_n (i),A)
\rangle \nonumber
\end{eqnarray}
with $A$ a generic state. Notice that in general the tensors
appearing in (\ref{tct}) and (\ref{atct}) will depend on the
coordinates $(t,{\bar t})$ of the space of theories. The logic
for for these rules is the equivalence between the insertion of a
field and the derivation with respect to the corresponding $t$
or $\bar t$ parameter. Considering that we want to study
the $t{\bar t}$ space and not the full phase space
available for the topological string, the derivation rule
associated to arbitrary gravitational descendants (\ref{d}.3)
should only involve their $t$-part.

\vspace{3mm}

\subsection{Consistency Conditions: Computations}

\vspace{2mm}

We pass now to study systematically
the consistency conditions for the contact term algebra defined
by equations (\ref{tct}) and (\ref{atct}). From the symmetry of
the operator-operator contact terms (\ref{csy}), the consistency
conditions
\begin{equation}
\int \! A\int \! B \; |C\rangle = \int \! B \int \! A \; |C \rangle
\end{equation}
for $A$, $B$, $C$ arbitrary operators, reduce to
\begin{equation}
\int \! A \: ( \! \int \! B \; |C\rangle) = \int \! B \: (\! \int \!
A \; |C\rangle)
\end{equation}
We will use from now on this simplified form.

Let us begin considering
\begin{equation}
\int \sigma_1 (i) \; (\!\int \! \sigma_1 (j) \; |k\rangle) =
\int \sigma_1 (j) \; (\!\int \! \sigma_1 (i) \; |k\rangle)
\label{1}
\end{equation}
{}From the contact term algebra (\ref{tct})-(\ref{atct}) and
the derivation rules (\ref{d}), we get
\begin{eqnarray}
\int \sigma_1 (i) (\! \int \! \sigma_1 (j) \; |k\rangle) =
\partial_i A_{jk}^l | \sigma_2 (l) \rangle + A_{jk}^n (A_{in}^l
|\sigma_2 (l) \rangle + C_{in}^l |\sigma_1 (l) \rangle ) + &&
\nonumber \\
+ \partial_i C_{jk}^l |\sigma_1 (l) \rangle + C_{jk}^n (
A_{in}^l | \sigma_1 (l) \rangle + C_{in}^l | l \rangle ) =
i \leftrightarrow j &&
\label{2}
\end{eqnarray}
Defining, according to (\ref{co}) and
(\ref{ct}), a covariant derivative by
\begin{equation}
D_i \equiv \partial_i - A_i
\label{4}
\end{equation}
we obtain from (\ref{2}) both the flatness condition
\begin{equation}
[D_i , D_j ] = 0
\label{3}
\end{equation}
and the integrability condition for the tensor $C_{ij}^k$
\begin{equation}
D_i C_{jk}^l = D_j C_{ik}^l
\label{5}
\end{equation}
where the connection $A$ acts only in the state indices $k$,
$l$. The connection associated to
the operator indices $i$, $j$ should be given by $c(i,j)$ which,
being symmetric, cancels from expression (\ref{5}).
It also follows from the consistency condition (\ref{2}) the
associativity of the tensor $C_{ij}^k$
\begin{equation}
C_{ik}^n C_{jn}^l = C_{jk}^n C_{in}^l
\label{6}
\end{equation}
Equations (\ref{5}) and (\ref{6}) imply that $C_{ij}^k$ are the
structure constants of the TFT. These equations, together with
(\ref{3}), are the $t$-part of the $t{\bar t}$-equations \cite{CV}.

We study next a consistency condition involving $t{\bar
t}$-contact terms
\begin{equation}
\int {\bar a} ( \!\int\! \sigma_1 (i) \;| j \rangle ) = \int
\sigma_1 (i) ( \!\int\! {\bar a} \;| j \rangle )
\label{7}
\end{equation}
{}From
\begin{eqnarray}
\int {\bar a} (\!\int\! \sigma_1 (i) \;| j \rangle ) & = &
\partial_{\bar
a} A_{ij}^k | \sigma_1 (k) \rangle + H_{{\bar a} k}^l ( A_{ij}^k
| \sigma_2 (l) \rangle + C_{ij}^k | \sigma_1 (l) \rangle) +
\partial_{\bar a} C_{ij}^k | k \rangle \nonumber \\
\int \sigma_1 (i) ( \!\int\! {\bar a} \;| j \rangle ) & = &
\partial_i
H_{{\bar a} j}^k | \sigma_2 (k) \rangle + H_{{\bar a} j}^n (
A_{in}^l | \sigma_2 (l) \rangle + C_{in}^l | \sigma_1 (l) \rangle)
\label{8}
\end{eqnarray}
we get a $t{\bar t}$-type equation for the connection $A_i$
\begin{equation}
\partial_{\bar a} A_{ij}^k =  [ H_{\bar a} , C_i ]_{j}^{k}
\label{9}
\end{equation}
the following constrain for the tensor $H_{\bar a}$
\begin{equation}
\partial_i H_{{\bar a} j}^k + A_{in}^k H_{{\bar a} j}^n -
A_{ij}^n H_{{\bar a} n}^k \equiv D_i H_{{\bar a} j}^k = 0
\label{10}
\end{equation}
and the holomorphicity of $C_{ij}^k$
\begin{equation}
\partial_{\bar a} C_{ij}^k = 0
\label{11}
\end{equation}
which is satisfied if $C_{ij}^k$ are the topological
structure constants.

The requirement
\begin{equation}
\int {\bar a} (\! \int \!{\bar b}\; |i \rangle ) = \int {\bar b} (\!
\int \!{\bar a} \;| i \rangle )
\label{12}
\end{equation}
implies more constrains on the tensor $H_{\bar a}$, namely
\begin{eqnarray}
\partial_{\bar a} H_{{\bar b} i}^k & = & \partial_{\bar b}
H_{{\bar a} i}^k  \label{13} \\
H_{{\bar a} i}^n H_{{\bar b} n}^k & = & H_{{\bar b} i}^n
H_{{\bar a} n}^k \nonumber
\end{eqnarray}

Before using (\ref{10}) and (\ref{13}) to solve $H_{\bar a}$, we
need another piece of information coming from the
contact term algebra.
In order to have a coherent interpretation of the contact term
\begin{equation}
\int P | {\bar a} \rangle = {\widetilde H}_{0 {\bar a}}^l | \sigma_1
(l) \rangle
\end{equation}
for $P$ the puncture operator, the simplest choice is to identify
the "formal" states $| {\bar a} \rangle$ with
\begin{equation}
| {\bar a} \rangle = M_{\bar a}^l | \sigma_2 (l) \rangle
\label{15}
\end{equation}
{}From this we immediately get
\begin{equation}
\int i | {\bar a} \rangle = \int i ( M_{\bar a}^l | \sigma_2 (l)
\rangle ) = ( \partial_i M_{\bar a}^l + M_{\bar a}^n A_{i n}^l )
| \sigma_2 (l) \rangle + C_{in}^l M_{\bar a}^n | \sigma_1 (l)
\rangle
\label{16}
\end{equation}
The consistency of (\ref{16}) with the contact term
\begin{equation}
\int i | {\bar a} \rangle = {\widetilde H}_{i {\bar a}}^l
| \sigma_1 (l) \rangle
\label{17}
\end{equation}
requires
\begin{eqnarray}
&& D_i M_{\bar a}^l  =  0  \label{18} \\
&& {\widetilde H}_{i {\bar a}}^l = C_{in}^l M_{\bar a}^n \nonumber
\end{eqnarray}
Assuming now that the matrix $M$ is invertible, we obtain from
equation (\ref{18}.1) that the
connection $A_i$ is given by
\begin{equation}
A_{ij}^k = ( \partial_i M_{j}^{\bar a} ) M_{\bar a}^k
\label{18a}
\end{equation}
where $M_{j}^{\bar a} M_{\bar a}^k \! = \! \delta_{j}^k$. The
matrix $M_{\bar a}^l$ can be understood as providing an
isomorphism between the topological and antitopological sectors.
Therefore, from eq. (\ref{15}), we can now interpret the formal
states $|{\bar a}\rangle$, introduced in (\ref{atct}.2), as a
gravitationally dressed version of the antitopological basis.

Using $M_{\bar a}^l$ as a topological-antitopological change
of basis, it is convenient to
redefine $H_{\bar a}$ in terms of a new tensor
${\bar C}_{{\bar a} {\bar b}}^{\bar c}$ as follows
\begin{equation}
H_{{\bar a} i}^l = {\bar C}_{{\bar a} {\bar b}}^{\bar c}
M_{i}^{\bar b} M_{\bar c}^l
\label{19}
\end{equation}
{}From (\ref{10}) and (\ref{13}), the tensor
${\bar C}_{{\bar a} {\bar b}}^{\bar c}$ should satisfy
\begin{eqnarray}
{\bar C}_{{\bar a} {\bar c}}^{\bar n} {\bar C}_{{\bar b}
{\bar n}}^{\bar d} & = & {\bar C}_{{\bar b} {\bar c}}^{\bar n}
{\bar C}_{{\bar a} {\bar n}}^{\bar d} \nonumber \\
D_{\bar a} {\bar C}_{{\bar b} {\bar c}}^{\bar d} & = &
D_{\bar b} {\bar C}_{{\bar a} {\bar c}}^{\bar d} \label{20} \\
\partial_i {\bar C}_{{\bar a} {\bar b}}^{\bar c} & = & 0 \nonumber
\end{eqnarray}
which are the defining relations for the structure constants of
an antitopological theory. Let us remark that (\ref{20}) does not
oblige the ${\bar C}_{\bar a}$ to be the structure constants of
the conjugate antitopological theory, i.e. ${\bar C}_{\bar a}
\!=\! (C_a)^{\ast}$. The only condition on the antitopological
theory is that its ring of observables has the same dimension as
the one of the topological theory with which it mixes.

Once we have fixed all the tensors appearing in the contact term
algebra, we can derive from (\ref{9}) and (\ref{19}),
the $t{\bar t}$-equation for the connection
\begin{equation}
\partial_{\bar a} A_{ij}^k = [ {\bar C}_{\bar a} , C_i ]_{j}^k
\end{equation}
Thus we have proved that
the complete set of $t{\bar t}$-equations appear as the
solution of the consistency conditions of the contact term
algebra (\ref{tct}), (\ref{atct}).

Before finishing this section, we would like to make some
comments. A very important ingredient in solving the consistency
conditions for the algebra (\ref{tct}) and (\ref{atct}),
was to suppose the symmetry of the
operator-operator contact terms (\ref{csy}). Indeed,
this is consistent with the solution we have obtained. Let us
consider the contact term between primary fields, according to
(\ref{cts})
\begin{equation}
c(i,j) \equiv A_{ij}^k \phi_k - A_{i0}^n C_{jn}^k \phi_k =
g_{0 {\bar b}} ( \partial_i C_{j {\bar a}}^{\bar b} ) g^{{\bar a}
k} \phi_k
\label{22}
\end{equation}
which is clearly symmetric in $i$, $j$ (see eq. (\ref{5})). This
extends trivially to the contact terms involving gravitational
descendants because the structure constants satisfy
\begin{equation}
C_{ij}^k = C_{ji}^k
\end{equation}

Based on these facts, we have assumed that $c({\bar
a},i)\!=\!c(i,{\bar a})$\footnote{Notice that the operator
involved in these operator-operator contact terms is the one
which should define the state $|{\bar a}\rangle$ given by
equation (\ref{15}). In fact the state-operator relation for the
antitopological sector involved in the contact term algebra
should formally be defined by (\ref{15}) and the condition
$c({\bar a},i)\!=\!c(i,{\bar a})$.}.

All the consistency conditions
are immediately satisfied by the solution presented,
with two exceptions
\begin{eqnarray}
\int i \int \sigma_n (j) | k \rangle & = & \int \sigma_n (j)
\int i | k \rangle \; \;, \; \; \; n>0 \label{24} \\
\int i \int {\bar a} | j \rangle & = & \int {\bar a} \int i
|j \rangle \nonumber
\end{eqnarray}
which both will involve factorization terms. The
necessity for including factorization terms at the level of
consistency conditions is already present in the simplest case
of contact term algebra, namely in pure topological gravity. In
that case, and due to the asymmetry of the contact algebra, it
is not possible to fulfill $\int {\hat P} \int {\hat \sigma}_n
|{\hat P} \rangle \! = \! \int {\hat \sigma}_n \int {\hat P}
| {\hat P} \rangle$\footnote{The asymmetry is originated by the
curvature factor that defines the corrected operators ${\hat
\sigma}_n \!=\! e^{\frac{2}{3}(n\!-\!1)\pi} \sigma_n$, where
$\pi$ is the conjugate of the Liouville field.} without
taking into account factorization terms.

A general contribution from a factorization term is
\begin{equation}
\int A \int B | C \rangle = {\cal B}_{B}^{\alpha \beta} C_{\beta
AC} | \alpha \rangle + ...
\label{25}
\end{equation}
where ${\cal B}_{B}^{\alpha \beta}$ is the factorization tensor,
the indices $\alpha$, $\beta$ run in principle over
gravitational descendants,
and $C_{\beta AC}$ are 3-point functions. The dots in (\ref{25})
mean the usual contact term contributions. By Witten's
recursion relations \cite{W1}, $C_{\beta AC}$ vanishes unless the
three indices are primary (see next section). Therefore the only
consistency
conditions in which factorization terms can appear are
(\ref{24}), as it is necessary.

The solution of (\ref{24}.1) involves a factorization tensor
satisfying
\begin{equation}
{\cal B}_{(n,j)}^{(n-2,l) r} C_{r ik} \equiv \partial_j C_{ik}^l
+ A_{jr}^l C_{ik}^r \;\;, \;\;\; n>0
\label{26}
\end{equation}
More interesting are the factorization terms contributing to
(\ref{24}.2), which are given by
\begin{equation}
{\cal B}_{\bar a}^{ij} = - H_{{\bar a} r}^i \eta^{rj} = -{\bar
C}_{\bar a}^{ij}
\label{27}
\end{equation}
where $\eta_{ij}$ is the topological metric. The consistency
conditions applied to a general string amplitude impose additional
restrictions on the factorization tensors, which in the case of
(\ref{27}) imply that the metric $\eta_{ij}$ is covariantly constant
\begin{equation}
D_k \eta_{ij} =0
\label{30}
\end{equation}
The hermitean metric $g_{i {\bar j}}$ defined in (\ref{i2}) can be
represented in the following way
\begin{equation}
g_{i{\bar j}} = \eta_{ik} M_{\bar j}^k
\label{31}
\end{equation}
in consequence, from (\ref{18}) and (\ref{30}), it is also
covariantly
constant. The existence on the space of theories of a topological
and a
hermitean compatible metrics, i.e. both of them covariantly
constants,
was showed in \cite{D} to be equivalent to the $t{\bar t}$-geometry.

The appearance of factorization terms for the $\bar a$ operators
can be motivated by the $n\!=\!2$ gravitational index
that they, heuristically, seems to carry as reflected in $|{\bar
a} \rangle \!=\! M_{\bar a}^l | \sigma_2 (l) \rangle$
(see eq.(\ref{15})). The main task of section 3 will be to
explore the consequences of the factorization term (\ref{27}) for
the $\bar a$ operator.
We will propose there a generalization of the holomorphic
anomaly equation, discovered in \cite{BCOV1} for ${\hat c}\!=\!3$
topological strings, to a generic topological string theory.

\vspace{3mm}
\subsection{Hodge Strings and Gravitational Descendants}

One of the main ingredients in the definition of string
amplitudes is the couple $(Q,b)$ with $Q$ the BRST operator and
$b$ the antighost. Their algebraic relations are given by:
\begin{eqnarray}
Q^2=b_{0}^2= 0 && \label{c1} \\
\{ Q, b_0 \} = H && \nonumber
\end{eqnarray}
which are formally similar to the ones defining a $N\! = \!2$
algebra. The main difference between (\ref{c1}) and the $N\!=\!2$
algebra is, as it was pointed out in \cite{BCOV1}, that for the
$N\!=\!2$ case the
cohomology of $b_{0}$ is non trivial and in fact isomorphic to
the BRST cohomology of $Q$, in contrast to what happens for the
bosonic string. By a Hodge string we mean one where
the $b_{0}$ antighost is defined by the $G_{0}$ component of the
SUSY current $G^-$:
\begin{equation}
G_0 = Q^{*} = \oint G^-
\end{equation}
with $Q^{*}$ the adjoint of the BRST charge. The name Hodge
comes from the fact that the couple $(Q,Q^{*})$ satisfies the
Hodge relations and in particular the $Q,Q^{*}$ lemma, i.e. any $Q$
closed form $a$ which is at the same time $Q^{*}$ exact can be
written as $QQ^{*}c$ for some $c$.

For the bosonic string theory the couple $(Q,b_{0})$ defines an
equivariant cohomology \cite{DN} by the conditions:
\begin{eqnarray}
Q|\chi\rangle & = & 0 \\
b_0 | \chi \rangle & = & 0 \nonumber
\end{eqnarray}
which characterizes the spectrum of physical states and the
invariance of string amplitudes with respect to changes of local
coordinates. The notion of equivariant cohomology is specially
relevant when we work with topological strings. In fact the
gravitational descendants appear as pure BRST states
$|\chi\rangle \!=\!Q |\psi \rangle$ non
trivial in the equivariant cohomology: $b_0 |\chi\rangle \!=\!0$,
$b_0 | \psi\rangle \! \neq \!0$ \cite{L,EY}. The existence of
these states is on the other hand crucial in the cancel
propagator argument computation of contact terms \cite{VV,DN,L}.

Let us define the contact term between two chiral fields as follows
\begin{equation}
\int_{0}^{\infty} \int_{0}^{2 \pi} d \tau d \phi e^{\tau T_+}
e^{\phi T_-} G_{0,+} G_{0,-} \phi_i (1) | \phi_j (0) \rangle
\label{c2}
\end{equation}
where we have used light type coordinates and where
$|\phi_j (0)\rangle$
represents the state defined at the boundary of the disc with
the field $\phi_j$ inserted at the origin. A non vanishing
contribution to this contact term will come from a $Q$ exact
part in the product $\phi_i \phi_j$, non trivial in the equivariant
cohomology defined with respect to $G_{0}$. We see in this way
that gravitational descendants and contact terms are two
different aspects of the same fundamental concept, namely the
one of equivariant cohomology. If in the definition (\ref{c2}) of
contact terms we use for $G_{0}$ the super-energy momentum
tensor, then, the so defined equivariant cohomology contains all
the gravitational descendants and we can symbolically represent
the contribution to the contact term as follows
\begin{eqnarray}
&& \phi_i \phi_j = \sigma_1 (\chi)+ ... \\
&& c(\phi_i, \phi_j)=\chi \nonumber
\label{c3}
\end{eqnarray}
These equations make clear the interplay between contact terms and
the
"matter representation" \cite{L,EY} of gravitational descendants.
In order to be precise it is important to make the following
remark concerning the contact term (\ref{c2}) for topological matter
coupled to topological gravity. The $G's$ appearing in (\ref{c2}),
which play in the construction of string amplitudes as measures
on the moduli space, a similar role to the $b_{0}$ antighost in
the bosonic string, are coming from the integration of the $N=2$
supermoduli and that is the reason they are defined by the
superpartner of the energy momentum tensor.
The situation changes dramatically if in equation (\ref{c2}) we use
for $G_{0}$ the Hodge pair of the BRST charge i.e we use the
susy current $Q^*$. In this case and as a simple consequence of
the $Q,Q^{*}$ lemma, any $Q$ exact state which is $Q^{*}$ closed
is trivial in the equivariant cohomology defined by the Hodge
pair $(Q,Q^{*})$ and therefore we can not any more interpret the
gravitational descendants as non trivial in the $(Q,Q^{*})$
equivariant cohomology. In a certain sense and in comparison with
the bosonic string, when we work in Hodge
strings we loose the richness of gravitational descendants in
the equivariant cohomology but we gain a non trivial cohomology
for the $b_{0}$ antighost. Thinking in these terms it seems
natural to expect that part of the physics which is ordinarily
associated with the presence of gravitational descendants will
appears in Hodge strings as a consequence of the non trivial
cohomology for the $b_{0}$ antighost. One nice example of this
phenomena is the holomorphic anomaly discovered in reference
\cite{BCOV,BCOV1} which crucially depends on the non trivial
cohomology of
the $b_{0}$ antighost and on the other hand looks formally very
similar to the recursion relations determined by gravitational
descendants. This interplay between the holomorphic anomaly and
gravitational recursion relations will become clear in the next
section in the context of the $t{\bar t}$-contact term algebra.

Why should we work with Hodge strings? In the philosophy
underlying this paper Hodge strings and therefore contact terms
defined with the $G_{0}$ part of the susy current $Q^*$, seems
the natural candidates to represent the variation of topological
matter amplitudes under an infinitesimal change of theory, or, in
other words, are the natural strings we should use to define the
$t\bar{t}$-geometry on the space of 2D-theories. A different
way to motivate the concept of Hodge strings is by introducing
the notion of covariantization used in ref \cite{BCOV1}.

In the Hodge case the integral representation (\ref{c2}) of the
contact
terms already give a good heuristic idea on the connection
between the non trivial cohomology of $b_{0}$ and the
$t\bar{t}$-geometry. In fact if we consider the derivative with
respect to $\bar t$ of (\ref{c2}), we will get a non trivial
contribution from the $\bar{t}$ dependence of $G_{0}$ which will
be absent in the case the cohomology of $G_{0}$ is trivial.
This contribution looks formally as the one expected
from the $t\bar{t}$-geometry, namely something like
\begin{equation}
\int_{0}^{\infty} \int_{0}^{2 \pi} d \tau d \phi e^{\tau T_+}
e^{\phi T_-} G_{0,+} Q_{0,+} {\bar \phi}_{\bar a} (1)
( \phi_i (1) | \phi_j (0) \rangle ) = {\bar \phi}_{\bar a}
(C_{ij}^k | k\rangle )
\label{c4}
\end{equation}
where the derivation of $G_{0,-}$ with respect to $\bar{t_{a}}$
have been replaced by an antichiral primary field
${\bar \phi}_{\bar a}$
which, after contracting with $C_{ij}^k |k\rangle$, will
produce the $t\bar{t}$-relation. Using a similar formal
argument we can give a functional integral interpretation of the
mixed $t\bar{t}$-contact terms we introduce in the previous
section. Thus we can interpret $|C({\bar a}, i)\rangle$ as
$\partial_{\bar a} |C(\sigma_1 , i)\rangle$, where again the
derivation
with respect to $\bar a$ is acting on $G_{0,-}$.

Concerning the dynamical meaning of gravitational descendants in
Hodge strings we can propose the following argument. If we insist
in defining the string equivariant cohomology in terms of the
Hodge pair $(Q,Q^{*})$, as it is the case in the computation of
the holomorphic anomaly in \cite{BCOV1}, then the corresponding
physical spectrum would be given by the harmonic zero energy
states. If now we want to compute a correlator involving
external gravitational descendant states the amplitude we will
get will fail to be invariant with respect to the peculiar type
of "reparametrizations" generated by $G_{0}$\footnote{Notice
that when we use for $G_{0}$ the Hodge pair of the BRST
charge $Q$ the condition on physical states to belong to the
corresponding equivariant cohomology is stronger than needed in
order to push down the string amplitude, as a measure on the
augmented moduli space of punctured Riemann surfaces with local
coordinates, to the moduli space of punctured Riemann surfaces.
This extra condition should be related to background
independence.}. This failure will
not affect the invariance with respect to changes of the local
coordinates on the world sheet but more likely
to "background" $\bar{t}$-independence,
the reason being the explicit dependence of $G_{0}$ on the
$\bar{F}$ part of the lagrangian\footnote{ Using the matter
representation of gravitational descendants in Landau-Ginzburg
theories (\ref{h8}) with $X$ the Landau-Ginzburg field, we observe
that $Q$
will contain the piece $W'dX$ and therefore $Q^{*}$ a piece like
$\bar{W}d\bar{X}$. Thus the failure with respect to the Hodge
equivariant condition will depends on the $\bar{t}$-couplings.}.
This argument
makes plausible to associate the $\bar{t}$ dependence of
covariant string amplitudes with the contribution of
gravitational descendants.

\vspace{7mm}

\section{The Holomorphic Anomaly}

\vspace{3mm}

For the case of $\hat{c}=3$ and for string amplitudes defined by
covariant derivatives of the partition function $F_{g}$, the
$t\bar{t}$-relations imply a $\bar{t}$ dependence of these
amplitudes which is known as the holomorphic anomaly. An
important ingredient in the derivation of the holomorphic
anomaly is the covariant definition of amplitudes in terms of
the $t\bar{t}$-connection. This is intimately related to our
previous discussion concerning Hogde strings. In fact the
covariantization of the amplitudes is forced when we want to
interpret them as determining the variation of topological
matter amplitudes on the space of couplings.
Technically this covariant definition of the string amplitudes
is easily done if there exists a vacuum line subbundle $\cal L$,
on the space
of couplings, such that the partition function $F_{g}$ is a
section of ${\cal L}^{2-2g}$. This can be achieved, in the
critical case $\hat{c}\!=\!3$, when we reduce the string amplitudes
to correlators between truly marginal fields and we identify the
space of couplings with the moduli space of the $\hat{c}\!=\!3$
$N\!=\!2$ super conformal field theory.

In this section we would like to propose a generalization of
the holomorphic anomaly to the general case where we do not
impose any restriction neither on the value of $\hat{c}$ or on
the type of string amplitudes. The logic for this generalization
is of course based on our representation of the
$t\bar{t}$-geometry in terms of stringy contact terms.
It is clear that if
we are not considering the critical $\hat{c}=3$ case we will
need to deal with amplitudes involving relevant fields and/or
gravitational descendants.

We will consider first the simplest case of genus zero amplitudes
$C_{i_1...i_s}$. In particular, let us begin by the first
non-trivial
case, namely, the 4-point function $C_{i_1 i_2 i_3 i_4}$ for chiral
primary fields. The 3-point function on the sphere is given by
\begin{equation}
C_{i_1 i_2 i_3} = C_{i_1 i_2}^n \eta_{n i_3}
\label{h1}
\end{equation}
where $i_1$ would corresponds to an operator index, while
$i_2$, $i_3$
are states indices. Motivated by this, we define
\begin{equation}
C_{i_1 i_2 i_3 i_4} \equiv D_{i_4} C_{i_1 i_2 i_3} =
\partial_{i_4} C_{i_1 i_2 i_3} - \Gamma_{i_4 i_1}^{n} C_{n i_2 i_3}
- A_{i_4 i_2}^n C_{i_1 n i_3} - A_{i_4 i_3}^n C_{i_1 i_2 n}
\label{h2}
\end{equation}
with $\Gamma_{i}$ and $A_{i}$ the operator-operator and
operator-state connections respectively
\begin{eqnarray}
&& \Gamma_{ij}^k = A_{ij}^k - A_{i0}^n C_{jn}^k = g_{0 \bar{a}}
(\partial_i C_{j \bar{b}}^{\bar a} ) g^{{\bar b} k} \\
&& A_{ij}^k = (\partial_i g_{j {\bar a}} ) g^{{\bar a} k} \nonumber
\label{h3}
\end{eqnarray}
Using that the topological metric $\eta_{ij}$ is covariantly
constant
and the integrability of the structure constants (\ref{5}), we get
that definition (\ref{h2}) is symmetric in all indices.

The $\bar t$-variation of the 4-point function,
$\partial_{\bar a} C_{i_1 i_2 i_3 i_4}$, has three type of
contributions.
{}From the first term on the r.h.s. of the $t{\bar t}$-equation
$\partial_{\bar a} A_{ij}^k \! = \!
[{\bar C}_{\bar a}, C_i ]_{j}^k$, we obtain
\begin{equation}
C_{i_1 i_2 n} {\bar C}_{\bar a}^{nm} C_{m i_3 i_4} + perm_{(i_1
i_2 i_3 i_4)}
\label{h4}
\end{equation}
which will be generically denoted from now on by
"$fact( {\bar C}_{\bar a})$".
The next contributions come from the
$t{\bar t}$-equations for both connections $A_i$ and $\Gamma_i$
\begin{equation}
- {\bar C}_{{\bar a} i_1}^n C_{n i_2 m} \eta^{mr} C_{r i_3 i_4} -
perm_{(i_1 i_2 i_3 i_4)}
\label{h5}
\end{equation}
It is important here the additional part $A_{i0}^n C_{jn}^k$ of the
connection $\Gamma_i$, because its $\bar t$-derivative
provides the term
$- {\bar C}_{{\bar a} i_4}^n C_{n i_1 m} \eta^{mr} C_{r i_2 i_3}$,
otherwise lacking. Finally, from the additional piece in
$\partial_{\bar a} \Gamma_i$ still
remains
\begin{equation}
{\bar C}_{{\bar a} 0}^n C_{n i_1 m} \eta^{mr} C_{r i_2 s}
\eta^{sl} C_{l i_3 i_4}
\label{h6}
\end{equation}

It is useful now to consider Witten's recursion relations
\cite{W1} for topological strings at genus zero
\begin{eqnarray}
&& \langle \sigma_{n_1} (i_1) \sigma_{n_2} (i_2) ... \sigma_{n-s}
(i_s) \rangle_0 = \label{h7} \\
&& \; \; \; \; = \sum_{X \cup Y=S} \langle \sigma_{n_1 - 1}
(i_1) \prod_{k \in X} \sigma_{n_k} (i_k) \: \alpha \:
\rangle_0 \: \eta^{\alpha \beta} \langle \: \beta  \prod_{l \in Y}
\sigma_{n_l} (i_l) \sigma_{n_{s-1}} (i_{s-1}) \sigma_{n_s} (i_s)
\rangle_0 \nonumber
\end{eqnarray}
where $S\!=\! \{i_2 ... i_{s-2} \}$ and $\alpha$, $\beta$ are
primary fields. For Landau-Ginzburg theories, in which it is
possible a matter representation of gravitational descendants,
they are given, restricted to the small phase space, by the
recursive formula \cite{EY}
\begin{equation}
\sigma_n (i) = W' \int^X \sigma_{n-1} (i) + \sum_{\alpha}
\langle \sigma_{n-1} (i) \alpha \rangle_0 \eta^{\alpha \beta}
\phi_{\beta}
\label{h8}
\end{equation}
with $W(X)$ the superpotential and $W'\!=\!\partial_X W$. Let us
consider the modified definition of gravitational descendants
\cite{L}
\begin{equation}
{\tilde \sigma}_n (i) = W' \int^X {\tilde \sigma}_{n-1} (i)
\label{h9}
\end{equation}
The operators ${\tilde \sigma}_n (i)$ satisfy analogous
recursion relations to (\ref{h7}), with the only difference that
there are no factorizations with less than three points, namely
the subset $X$ should contain at least one point, and all
3-point functions involving a field ${\tilde \sigma}_n (i)$,
$n>0$, vanish. We recall that this last condition was needed
in solving the consistency conditions (see paragraph after
(\ref{25})). In consequence, the gravitational descendants appearing
in this paper corresponds to definition (\ref{h9}), instead of
(\ref{h8}). We will drop from now on the tilde in
${\tilde \sigma}_n (i)$ to simplify notation.

Using the recursion relations for operators
(\ref{h9})\footnote{We are assuming here that Witten's recursion
relations are valid for covariantized  string amplitudes.}, it
is easily seen that (\ref{h5}) and (\ref{h6}) are associated to
$\sigma_1 (i)$ and $\sigma_2 (i)$ contributions respectively.
Therefore, collecting (\ref{h4})-(\ref{h6}), we obtain that the
$\bar t$-variation of the 4-point function can be written in the
following way
\begin{equation}
\partial_{\bar a} C_{i_1 i_2 i_3 i_4} = fact({\bar C}_{\bar a}) -
\sum_{i=1}^n {\bar C}_{{\bar a} i}^l C_{ \sigma_1 (l) i_1 ..
{\hat i} .. i_4} + {\bar C}_{{\bar a} 0}^l C_{\sigma_2
(l) i_1 i_2 i_3 i_4}
\label{h10}
\end{equation}

The existence of a non-vanishing $\bar t$-derivative relies in
covariantization. Indeed, if we work with non-covariant
genus zero amplitudes
\begin{equation}
C_{i_1 ... i_s} \equiv \partial_{i_s} ... \partial_{i_4} C_{i_1
i_2 i_3}
\label{h11}
\end{equation}
the $\bar t$-variations are zero due to the holomorphicity of
the 3-point function and the commutativity of the $t$ and $\bar
t$ partial derivatives, $[ \partial_i , \partial_{\bar a} ]\!=\!0$.

According to this, we can try to define covariant string
amplitudes by requiring that the $\bar t$-derivatives are given
by the generalization of (\ref{h10})
\begin{equation}
\partial_{\bar a} C_{i_1 ... i_s} = fact({\bar C}_{\bar a}) -
\sum_{i=1}^s {\bar C}_{{\bar a} i}^l C_{ \sigma_1 (l) i_1 ..
{\hat i} .. i_s} + {\bar C}_{{\bar a} 0}^l C_{\sigma_2
(l) i_1 ... i_s}
\label{h12}
\end{equation}
This expression is symmetric in all the indices by induction
because the 3- and 4-point functions are, as it is required.

The holomorphic anomaly equation (\ref{h12}) for generic
topological strings can be understood as following from the
contact term algebra. In fact, the first contribution to
(\ref{h12}) comes from the factorization tensor for the
antitopological operators $\bar a$ introduced in (\ref{27}). It
is in the second and third terms where resides the main
difference with the critical case. The second comes from the
mixed contact terms (\ref{atct}), and the third can be
interpreted as a bulk contribution, a priori allowed by the
symmetric contact term algebra (\ref{csy}) we are working with.
Notice that it is licit to give a meaning to covariant
derivatives in a purely stringy way, without making an extra
assumption on the existence of a vacuum line subbundle, thanks
to the interpretation, worked in section 2.2, of $t{\bar
t}$-connection as contact terms.

An important property of the last term in (\ref{h12}) is that,
due to charge conservation, it will cancel when we restrict
ourselves to a moduli $(t,{\bar t})$-point and marginal
perturbations. Indeed, expression (\ref{h12}) reduces to the
holomorphic anomaly equation presented in \cite{BCOV1} for the
critical ${\hat c}\!=\!3$ string and moduli perturbations
\begin{equation}
\partial_{\bar a} C_{i_1 ... i_s}^g = fact ( {\bar C}_{\bar a} )
- \sum_{i=1}^s G_{i {\bar a}}
(2g\!-\!2\!+\!s\!-\!1) C_{i_1 .. {\hat i} .. i_s}^g
\label{h13}
\end{equation}
where the second contribution in the r.h.s. is given in terms of
the Zamolodchikov metric $G_{i{\bar a}}$ \cite{Z},
an object which
requires for its definition the existence of a well defined
vacuum line subbundle, something which is not possible in the
generic case we are considering.
The equivalence of expressions (\ref{h13}) and (\ref{h12}) for a
moduli
$(t,{\bar t})$-point and marginal perturbations $i_j$ and $\bar a$,
is achieved because in this particular case the Zamolodchikov
metric
can be represented as \cite{CV}
\begin{equation}
{\bar C}_{{\bar a} i}^n = \frac{g_{i {\bar a}}}{g_{0{\bar 0}}}
\delta_{0}^n = G_{i {\bar a}} \delta_{0}^n \; ,
\end{equation}
and the factor
$(2g\!-\!2\!+\!s\!-\!1)$, which is the curvature of a
Riemann surface of genus $g$ and $s\!-\!1$ punctures,
corresponds to a dilaton insertion $C_{\sigma_1 i_1 .. {\hat
i} .. i_s}$.

It is also important to stress the similarity between the
two first contributions to the holomorphic anomaly once we use
Witten's recursion relations for gravitational descendants. Again
this makes clear the strong interplay between the way the $\bar
a$ field is factorizing the surface (the first term in
(\ref{h12})), something that in the original derivation of the
holomorphic anomaly \cite{BCOV1} is due to the non-trivial
cohomology of the $b_0$ antighost in Hodge (covariant) amplitudes,
and the factorization rules for gravitational descendants (the
second term in (\ref{h12})).

Based on the previous discussion, it seems plausible to
conjecture that the holomorphic anomaly equation obtained is
valid for any genus
\begin{equation}
\partial_{\bar a} C_{i_1 ... i_s}^g = fact_{g} ({\bar C}_{\bar a}) -
\sum_{i=1}^s {\bar C}_{{\bar a} i}^l C_{\sigma_1 (l) i_1 ..
{\hat i} .. i_s}^g + {\bar C}_{{\bar a} 0}^l C_{\sigma_2
(l) i_1 ... i_s}^g
\label{h14}
\end{equation}
where
\begin{equation}
fact_g ({\bar C}_{\bar a}) = \frac{1}{2} {\bar C}_{\bar
a}^{\alpha \beta} C_{\alpha \beta \: i_1 ... i_s}^{g-1} +
\frac{1}{2} {\bar C}_{\bar a}^{\alpha \beta}
\sum_{r=0}^g \sum_{X \cup Y =S} C_{\alpha \: j_1 ... j_l}^r
C_{\beta \: j_{l+1} ... j_s}^{g-r}
\end{equation}
with $X\!=\!\{j_1 ... j_l\}$ and $Y\!=\!\{j_{l+1}...j_s\}$.
As an indication of the validity of (\ref{h14}) we can analyze
the genus one case, in which recursion relations analogous to
(\ref{h7}) hold \cite{W1}. It is straightforward to see that if
$C_{i}^1\!\equiv\! \partial_i F^1$ satisfies (\ref{h14}), then
$C_{ij}^1 \! \equiv \! D_i F_{j}^1 \!= \! (\partial_i -
\Gamma_i) F_{j}^1$ also does.

\vspace{3mm}

{\bf Acknowledgments}

We would like to thank I. Krichever and A. Losev for many
valuable discussions.
This work was partially supported by
european community grant ERBCHRXCT920069,
and by grant PB 92-1092, and the work
of E.L. by M.E.C. fellowship AP9134090983.

\end{document}